\documentclass[letter]{aa} 

\usepackage{graphicx}
\usepackage{txfonts}
\begin{document}

   \title{The heavy-elements heritage of the falling sky}


   \author{Alejandra Recio-Blanco
          \inst{1}
          \and
          Emma Fern\'andez-Alvar\inst{1}
          \and
          Patrick de Laverny\inst{1}
          \and
          Teresa Antoja\inst{2} 
          \and \\
          Amina  Helmi\inst{3}
          \and
          Aur\'elien Crida\inst{1}
          }

   \institute{Universit\'e C\^ote d'Azur, Observatoire de la C\^ote d'Azur, CNRS, Laboratoire Lagrange, France \\
              \email{arecio@oca.eu}
         \and Institut de Ci\`encies del Cosmos, Universitat de Barcelona (IEEC-UB), Barcelona, Spain
           \and Kapteyn Astronomical Institute, University of Groningen, Landleven 12, 9747 AD Groningen, The Netherlands
             }

   \date{Received ...; accepted ...}

 
  \abstract
   {A fundamental element of galaxy formation is the accretion of mass through mergers of satellites or gas. Recent dynamical analysis based on Gaia data have revealed major accretion events in Milky Way's history. Nevertheless, our understanding of the primordial Galaxy is hindered because the bona fide identification of the most metal-poor and correspondently oldest accreted stars remains challenging.}
   {Galactic Archaeology  needs a new accretion diagnostic to understand primordial stellar
populations. 
Contrary to $\alpha$-elements, neutron-capture elements
present unexplained large abundance spreads for low metallicity stars, that could result from a mixture of formation sites.  }
   {We have analysed the abundances of yttrium, europium, magnesium and iron in Milky Way satellite galaxies, field halo stars and globular clusters.
   The chemical information has been complemented with orbital parameters based on Gaia data. In particular, the orbit's average
inclination has been considered. }
   {The [Y/Eu]  abundance behaviour with respect to the [Mg/Fe] turnovers for satellite galaxies of different masses reveals that higher luminosity systems, for which the [Mg/Fe] abundance declines at higher metallicities, present enhanced [Y/Eu] abundances, particularly in the [Fe/H] regime between -2.25~dex and -1.25~dex.
   In addition, the analysis has uncovered a chemo-dynamical correlation for both globular clusters and field stars of the Galactic halo,
   accounting for about half of the [Y/Eu] abundance spread. In particular,  [Y/Eu] under-abundances typical of protracted chemical evolutions, are preferentially observed in polar-like orbits, pointing to a possible anisotropy in the accretion processes.  }
   {Our results strongly suggest that the observed [Y/Eu] abundance spread  in the Milky Way halo could result from a mixture of systems with different masses.
   They also highlight that both nature and nurture are relevant to the Milky Way's formation, since its primordial epochs, opening new pathways for chemical diagnostics of our Galaxy building up.}


   \maketitle
%

\section{Introduction}
 The most primitive Galactic stars are essential to understand 
the Milky Way formation. Nevertheless, the robust identification of accreted 
objects is particularly challenging for stars with primordial abundances
having at most 30 times less metals than the Sun ([Fe/H]$\lesssim$-1.5). 
Kinematical or dynamical indications of accretion 
are insufficient  to reveal ancient mergers \citep[][]{JeanBaptiste17}.
They need to be 
complemented by chemical diagnostics \citep[][]{FreemanJoss}, 
as the chemical evolution of a system strongly depends on its mass. Compared to the massive
Milky Way, satellite galaxies 
generally present signs of protracted evolutions, being more metal deficient
and showing a variety of chemical patterns that we should retrieve in the accreted populations,
now mixed with in situ formed stars.

The most commonly used chemical diagnostic of accretion is the $\alpha$-elements (O, Mg, Si, S, Ca, Ti)  ratio with respect to iron ([$\alpha$/Fe]). Initially enhanced, 
the [$\alpha$/Fe] abundance starts to strongly decline with metallicity after the supernovae Ia explosion rate reaches a maximum \citep[][]{MatteucciGreggio}.
This produces a knee in the [$\alpha$/Fe] vs. [Fe/H] trend whose location provides constraints on the system total mass: the less massive
the system, the more metal-poor is the [$\alpha$/Fe] turnover. Unfortunately, this accretion diagnostic is not discriminating enough
for stars  in the Galactic halo, with metallicities lower than the  [$\alpha$/Fe] turnover of most satellite galaxies. As a consequence, metal-poor field stars 
kinematically proposed to be members of ancient accreted satellites, like Gaia-Enceladus/Sausage \citep[][]{AminaEnceladus,Sausage}, have similar
[$\alpha$/Fe] abundances as non-members for [Fe/H]$\lesssim$-1.5~dex. They only appear as a separate
sequence at higher metallicity \citep[][]{AminaEnceladus},  hampering also the detection of low mass mergers.
Similarly, the population of clusters in the Galactic halo is mostly homogeneous in their [$\alpha$/Fe] abundances \citep[][]{DualGalaxy18}. 

Galactic Archaeology thus needs a new accretion diagnostic to understand the primordial stellar
populations and, in this work, we have used neutron-capture elements to identify it.  
Contrary to $\alpha$-elements, neutron-capture elements
present unexplained large abundance spreads for low metallicity stars, that could result from a mixture of formation sites. In particular,
we have considered the logarithm of the ratio of a star's yttrium abundance with respect to its europium one, [Y/Eu].   Approximately 75\% of the solar
Yttrium was produced \citep[][]{Nikos18}  by low and intermediate mass asymptotic giant branch (AGB) stars, through slow neutron captures (relatively to the $\beta$-decay rates of unstable nuclei). In addition, first peak s-elements like Y have a larger contribution from low mass stars than second peak elements like Ba.  On the other hand, 94\% of europium is produced by massive stars through rapid neutron captures \citep[][]{Bisterzo14}. Proposed Eu production sites are neutron star mergers \citep[][]{Rosswog99}, high energy winds accompanying core collapse supernovae explosions \citep[][]{Woosley94} or magneto-hydrodynamical explosions of fast rotating stars \citep[][]{Winteler12}.  As a consequence, the [Y/Eu] abundance ratio characterizes the relative contribution 
of low-intermediate mass stars with respect to high mass stars, being therefore a good indicator of the chemical evolution efficiency.

\section{Chemical abundances and orbital parameter estimations}
The present study relies on several samples of objects: globular clusters
and field stars, both from the Milky Way and its satellites. We have made use of abundances of europium, yttrium and [Mg/Fe], collected from different literature works (c.f. Table~1 and further details in the Appendix). 
Concerning the [Y/Eu] uncertainty estimates, we have examined the abundance dispersion of the objects analysed by more than one study, including the dwarf stars database. 
The mean dispersion in the [Y/Eu] ratio is 0.07 dex, indicating a reasonable agreement between different literature sources. To adopt a conservative value, we have multiplied that dispersion by 2,  adopting  a typical error-bar of 0.15~dex. 
\smallskip

The chemical analysis of Milky Way objects has been complemented with orbital parameters based on Gaia data \citep[][]{GaiaDR2}. 
For globular clusters,  the orbital parameters are taken from Model-2 in \cite{AminaGaiaDR2}. 
They have been  computed as the average values over 10 Gyr of integration. 
To this purpose, we used the median values obtained from 1000 orbits for each cluster obtained through Monte Carlo realizations of the initial conditions, considering the observational measurements 
and their errors. In particular, the orbit's average inclination has been computed as arccos(Lz/Ltot). In our convention,
the orbital inclination is defined from the Galactic plane and comprised between 0$^{\circ}$ and 180$^{\circ}$, with prograde orbits
below 90$^{\circ}$. 
Error bars in the orbital parameters associated to model assumptions, have been estimated by comparing the results obtained with different Galactic potentials \citep[defined as Model-1, -2, -3 in][]{AminaGaiaDR2}. In particular, the dispersion in the orbital inclination (estimated as the third quantile value of the differences distribution between two models) is 6 degrees. 
In addition to this main dataset of cluster orbits, we have completed the sample with six additional objects from \cite{Vasiliev19}.


For our field stars samples, we have derived  the orbital parameters  using the python package galpy \citep[][]{galpy15}. We assume the MWPotential14 Milky Way mass model included in this package. 
We derived the action parameters through the action-angle isochrone approximation \citep[][]{galpy14}. 
As input parameters we have used the radial velocities gathered in Simbad, the Gaia DR2 proper motions and the distances from \cite{CorynDist18}.  In addition, we have checked the effect of using two different methodologies of the dynamical parameters for clusters and field stars. To this purpose, we have re-computed the clusters orbital inclinations using the field stars methodology
calculated the differences with respect to the Model-2 orbital results from Gaia Collaboration et al. 2018. The median absolute deviation of the orbital inclination differences is 2.5 degrees, confirming the consistency of the two approaches. 

Finally, we have assessed the impact of the detected Gaia kinematic biases \citep[][]{ShoenrichBias} in our data. For the field star samples, only 18 targets 
had a few parameters outside the Schoenrich et al. quality cuts and were excluded from the analysis. Regarding the globular cluster data, the \cite{AminaGaiaDR2} database is within the quality cuts, and the Vasiliev et al. compilation uses literature distances
and line-of-sight velocities not concerned by the Gaia parallax bias.

\section {Chemo-dynamical correlations and abundance spread in the Halo}
\begin{figure}[ht]
\centering
\includegraphics[width=9.5cm,height=6cm]{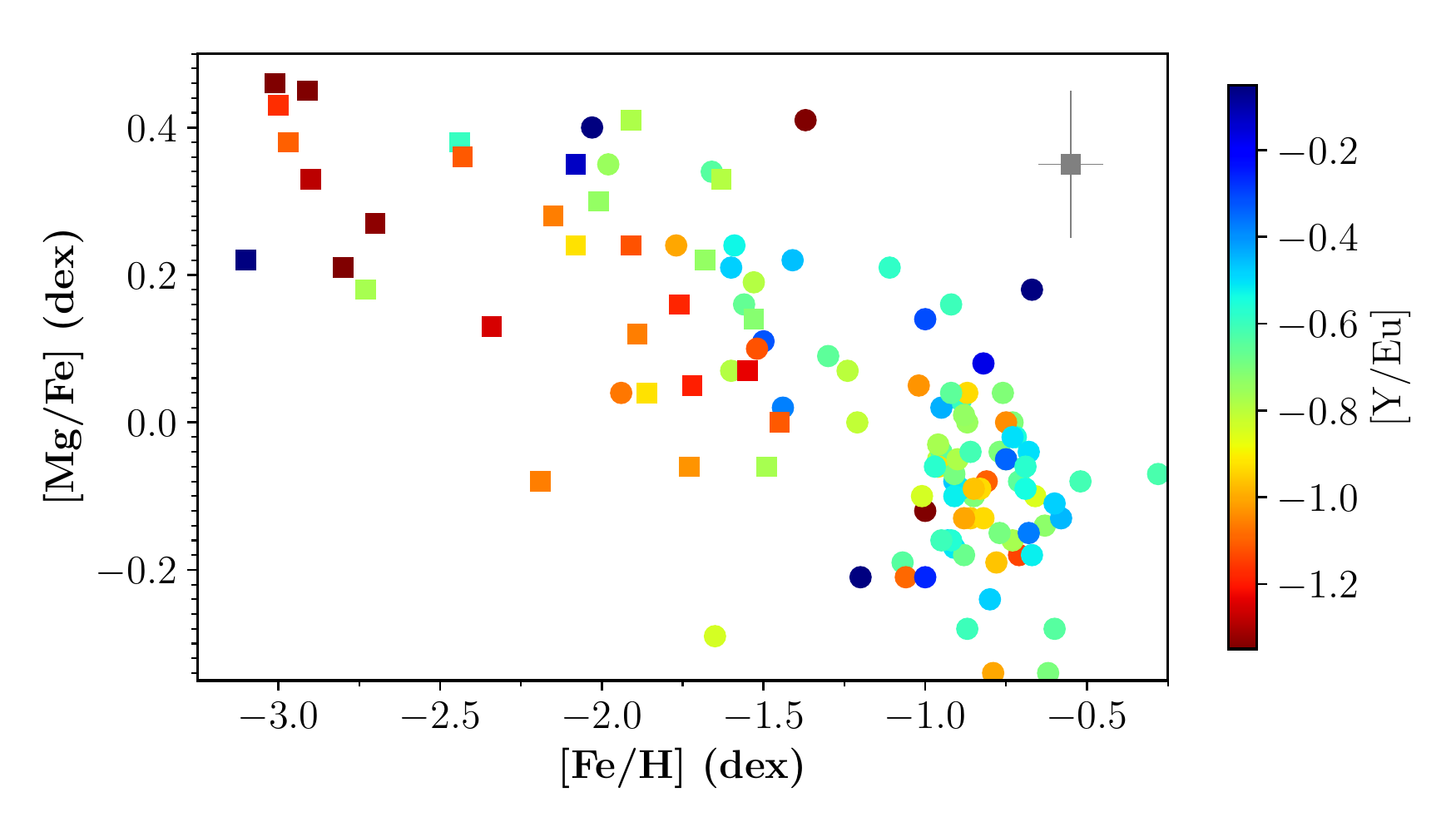}
\caption{Mg abundance with respect to iron as a function of [Fe/H] for stars belonging to
low mass satellites (Ursa Minor, Draco and Carina; square symbols) and to higher mass satellites (Fornax, Sculptor and Leo I; circles).
Points are colour coded by the stars [Y/Eu] content. }
\label{}
\end{figure}

 First of all, we have analysed the [Y/Eu]  behaviour with respect to the [Mg/Fe] turnovers for satellite galaxies of
different masses. Figure~1 shows the Mg abundance (an $\alpha$-element)  with respect to iron as a function of [Fe/H] for stars belonging to
low mass satellites (Ursa Minor, Draco and Carina; square symbols) and to higher mass satellites (Fornax, Sculptor and Leo I; circles).
The points are colour coded by the stars [Y/Eu] content. It can be observed that  higher luminosity systems, for which the [Mg/Fe]
abundance declines at higher metallicities, present enhanced [Y/Eu] abundances, particularly in the [Fe/H] regime between
-2.25~dex and -1.25~dex (see the Appendix for a separate [Y/Fe] and [Eu/Fe] analysis). 

\smallskip

Following the previous result, the observed [Y/Eu] abundance spread  in our Milky Way  could result from a mixture of systems with different masses.
If this is the case, the [Y/Eu] indicator should be compatible with the commonly used [Mg/Fe] accretion diagnostic, also in our Galaxy. 
This has already been observed in the high metallicity regime \citep[][]{Fishlock17},
but it is difficult to test in the metal-poor one, where  the [$\alpha/$Fe]  spread is very low.

\begin{flushleft}
\begin{figure*}[ht]
\includegraphics[width=17.5cm,height=9cm]{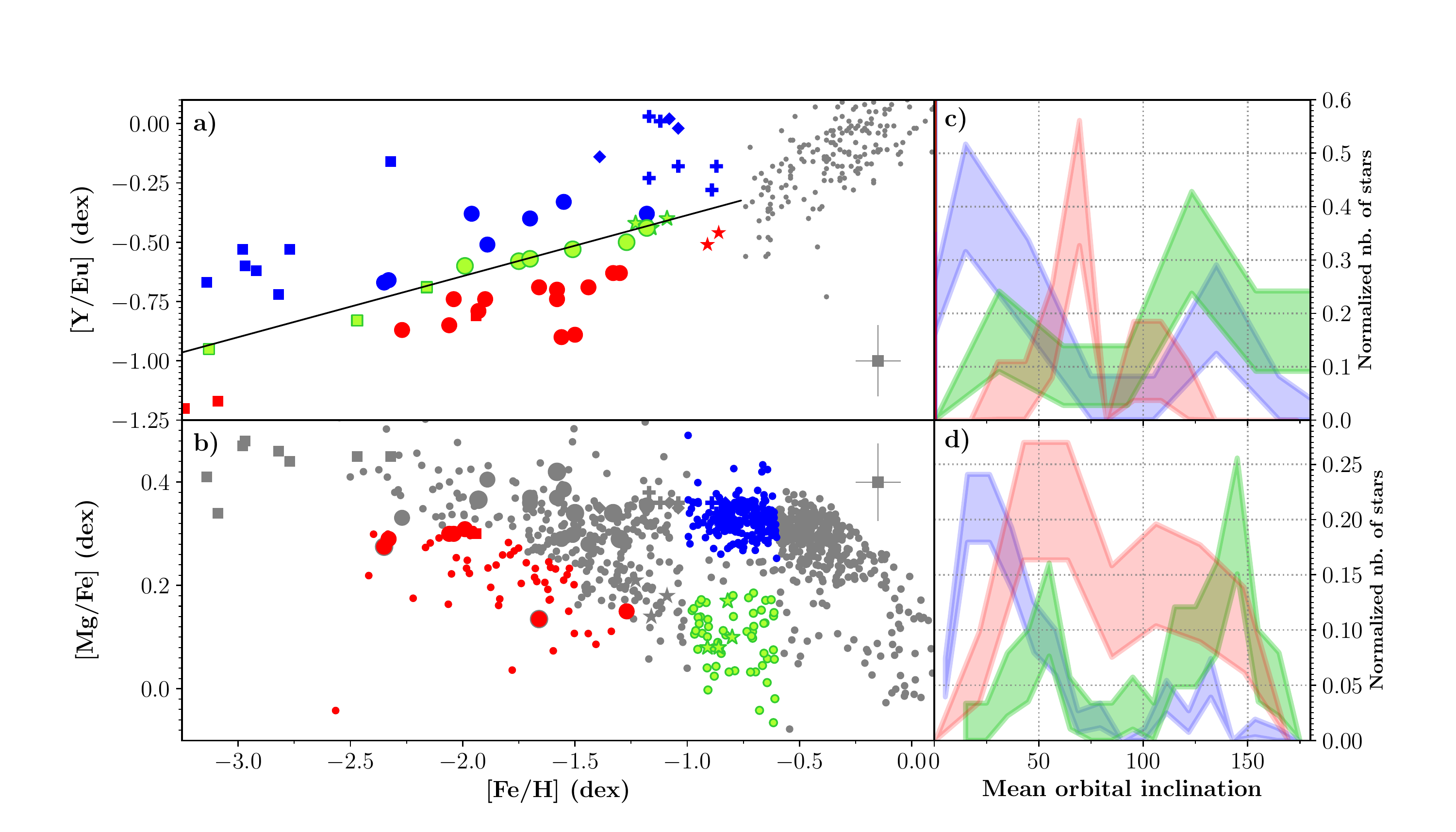}
\caption{Panel a:  Yttrium abundance with respect to europium as a function of iron abundance for Milky Way globular clusters (large circles) and field stars 
(squares for a metal-poor sample \citep[][]{Roederer14}, diamonds, stars and crosses for an intermediate-metallicity compilation \citep[][]{Fishlock17} of high-[Mg/Fe], low-[Mg/Fe] and thick disc stars respectively). The colour code is based on the [Y/Eu] departures from the standard value (black line). 
Panel b: [Mg/Fe] abundance ratio with respect to iron for the previous objects, when available, and a compilation of high transversal velocity stars from APOGEE. The colour code selects groups with different [Mg/Fe] turnovers (thus parent system masses).
Panels c and d:  distributions of orbital inclinations for the groups selected with the [Y/Eu] and the [Mg/Fe] criteria, respectively.
}
\label{}
\end{figure*}
\end{flushleft}

\begin{center}
\begin{figure}[ht]
\centering
\includegraphics[width=9.5cm,height=9cm]{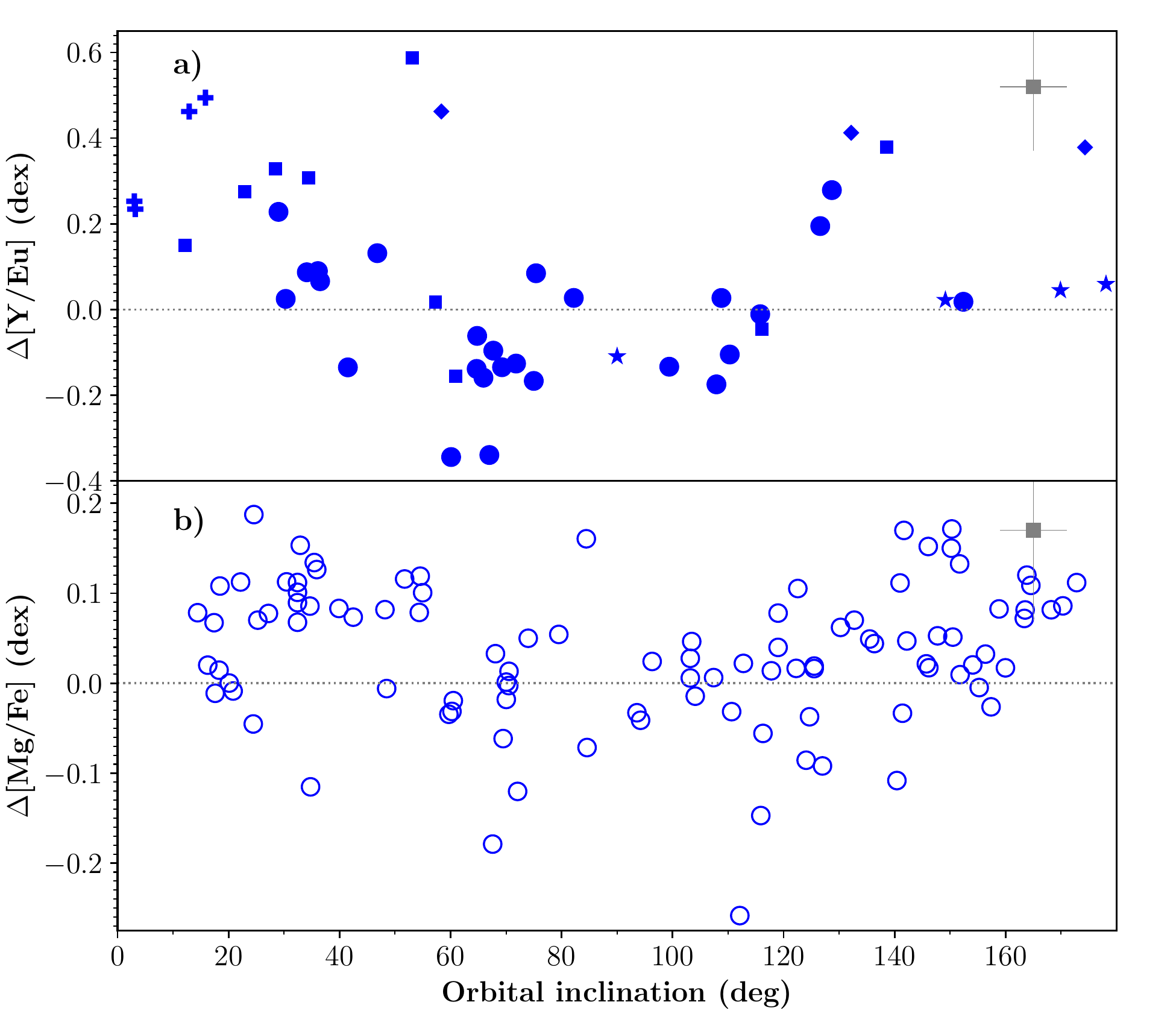}
\caption{Deviations in [Y/Eu] (panel a) and  [Mg/Fe] (panel b) abundances, with respect to the average, as a function of the orbital inclination.
No objects in common to the [Y/Eu] and the [Mg/Fe] analysis are included. Average values have been defined by a Theil-Sen 
linear fit for each abundance trend with metallicity. The [Mg/Fe] analysis is restricted to -2.0 $\le$ [M/H] $\le$-1.2 dex, to reduce
the non linear effect of the [M/Fe] turnover.}
\label{}
\end{figure}
\end{center}

\smallskip

Fortunately, since the arrival of precise Gaia astrometric
data, dynamical information can be used to break down this degeneracy. Indeed, chemo-dynamical correlations
retrieved both in the [Mg/Fe]  and the [Y/Eu] spread  could reinforce the  [Y/Eu] abundance as a good accretion indicator.
To test this possibility, our  Milky Way objects have been classified into three categories,  using the [Y/Eu]  and the [Mg/Fe] criteria 
independently (upper and lower panels of Figure~2, respectively):
first, objects with depleted [Y/Eu] values or metal-poor 
[Mg/Fe] turnovers (red targets) compatible with low-mass progenitors; second, objects  with intermediate [Y/Eu] abundances or
intermediate-metallicity [Mg/Fe] turnovers (green targets) possibly formed in higher-mass systems; and third, targets with
enhanced [Y/Eu] values or a metal-rich [Mg/Fe] turnover typical of the Milky Way in situ population (blue objects).
The [Mg/Fe]-selected samples act here as control groups testing the [Y/Eu] diagnostic.

Panels {\it c} and {\it d} show the normalized distribution of orbital inclinations for the three sets of objects, selected either with the [Y/Eu] diagnostic
or with the [Mg/Fe] one, respectively. Although the two chemical diagnostics target different objects (those in common being excluded from panel d
histograms) and span different metallicity regimes, the similarities between panels {\it c} and {\it d} 
distributions are important. Two-sampled Kolmogorov-Smirnov tests between the nine possible pairs of distributions have been performed to test this
similarity.
The null hypothesis, assuming that the samples come from a population with the same distribution,  is rejected for all the pairs except those having the same colour (targeting therefore the same parent system mass). 
In particular, depleted [Y/Eu] objects tend to present high orbital inclinations, 
as targets with a metal-poor [Mg/Fe] turnover. On the contrary, objects with intermediate [Y/Eu] abundances
and intermediate metallicity  [Mg/Fe] turnovers display mainly low inclination retrograde orbits. Finally, 
targets with high  [Y/Eu]  ratios and metal-rich [Mg/Fe] turnovers  show primarily
low inclination prograde orbits. As expected, adjacent groups in [Y/Eu] or [Mg/Fe] abundances (red-green and green-blue pairs), partially
overlap in their orbital inclination distributions as a result of  abundance uncertainties, but also to the fact that no
perfectly separated components seem to exist. In particular, in situ formed objects dynamically heated by past mergers \citep[e.g.][]{Belokurov19, Paola19} could also blur the orbital inclination distributions.

The above result confirms the coherence of the [Y/Eu] diagnostic with the [Mg/Fe] one, revealing possible chemo-dynamical correlations
with two independent chemical indicators. 
To quantify those trends, Figure~3 shows the deviations in [Y/Eu] and  [Mg/Fe] abundances with respect to the average,  as a function of orbital inclination.
Contrary to the analysis of Figure~2, no data subsamples are predefined and the considered metallicity regime spans  -2.0 $\le$ [M/H] $\le$-1.2 dex in both panels.
The two chemical diagnostics show under-abundances around the polar direction (60$^{\circ}$ $\lesssim$ inclination  $\lesssim$ 120$^{\circ}$)
and over-abundances  near the plane (prograde objects with inclination $\lesssim$ 60$^{\circ}$ and
retrograde objects with inclination $\gtrsim$120$^{\circ}$).   The observed 
chemo-dynamical correlations, including both globular clusters and field stars, are more pronounced for the [Y/Eu] abundances than for the [Mg/Fe] ones as expected from their corresponding
abundance spreads in this metallicity regime. In particular, the orbital inclination seems to account for about half of the [Y/Eu] abundance scatter.
   
\section{Conclusions}
Although Galactic studies need to be constantly validated in the huge parameter space of
Milky Way populations, the observed chemo-dynamical correlations open new paths of exploration 
of our Galaxy formation history. 
In the light of the previous conclusions, the heavy elements abundance scatter of the primordial Milky Way possibly results from
an amalgam of systems with different masses and chemical evolutions. 

First, objects in polar-like orbits showing underabundances of [Y/Eu] could result from a composite debris from
low mass accretions. Interestingly, polar orbits are also found for more recent merger events as the Sagittarius one. This suggests the possible existence 
of a preferential accretion axis around the polar direction, linking the Milky Way to its satellites and deserving further study.
In the metal-poor and intermediate metallicity regime, where the [Y/Eu] under-abundances are larger
than the [$\alpha$/Fe] ones, future large scale heavy-element studies seem crucial to distinguish between low-mass accretions and
slow rotating debris from more massive mergers.


Second, satellite merger debris in  retrograde orbits was previously suggested by the analysis of several dynamical overdensities
\citep[e.g.][]{AminaEnceladus, Sausage, Myeong19}, and attributed to high mass progenitors (Gaia Enceladus/Saussage, Sequoia).
In our study, the chemical patterns dominating that retrograde regime near the plane are indeed typical of high mass systems, reaching metallicities of -0.5~dex and relatively high [Y/Eu] abundances. The interplay of this old retrograde population with the prograde disc and
the slow rotating accretion debris is probably an important piece of the Galaxy formation puzzle.

Third, a prograde population, showing [Y/Eu] overabundances, seems to be present even in the low metallicity regime. 
It could be the fossil signature of the primitive collapsed Galaxy, probably occupying
prograde orbits near the plane, as the more metal-rich disc. This hypothesis is strengthen by the recent discovery of
very metal-poor stars with disc like orbits \citep[][]{Sestito20}

In conclusion, both nature and nurture appear to have played a role to build up the ancient Milky Way, leaving inprints we are starting to decode.
Chemical diagnostics, including heavy elements abundances, will certainly be fundamental in the on going Gaia revolution.   
   

\begin{acknowledgements}
This work has made use of data from the European Space Agency (ESA) mission Gaia Data Processing and Analysis Consortium
(https://www.cosmos.esa.int/gaia), processed by the Gaia Data Processing and Analysis Consortium.
(DPAC, https://www.cosmos.esa.int/web/gaia/dpac/consortium). Funding for the
DPAC has been provided by national institutions, in particular the
institutions participating in the Gaia Multilateral Agreement. ARB,
PdL and EFA acknowledge financial support from the ANR 14-CE33-014-01.
TA has received funding from the European Union's Horizon 2020
research and innovation programme under Marie Sklodowska-Curie grant
agreement number 745617 and also acknowledges funding from the MINECO
(Spanish Ministry of Economy) through grants ESP2016-80079-C2-1-R
(MINECO/FEDER, UE) and ESP2014-55996-C2-1-R (MINECO/FEDER, UE). AH
acknowledges funding from a Vici grant from the Netherlands
Organisation for Scientific Research (NWO).  We thank E. Vasiliev for providing
his orbital parameters for globular clusters.
ARB thanks Vanessa Hill,
Sebastian Peirani and Oliver Hahn for useful discussions and Chris Wegg for kindly language corrections.

\end{acknowledgements}

\bibliographystyle{aa} 
\bibliography{biblio}

\begin{appendix}
\section{Complementary information on literature abundances}

The adopted references for the abundances of the different elements and populations analysed  in this work are summarized in 
Table 1.A. The study of globular clusters chemical abundances is currently confined to heterogeneous compilations from different groups.
Nevertheless, despite these words of caution, clusters benefit today from several decades of efforts in chemical abundance estimations. 
The analysed Milky Way field stars abundances come from three different compilations: a photometric selection of metal-poor stars \citep[][]{Roederer14}, 
a study of heavy-element abundances for high-$\alpha$ and low-$\alpha$ stars at intermediate metallicity \citep[][]{Fishlock17} and a selection of high transversal
velocity stars from the APOGEE survey \citep[][]{ApogeeDR14}.
When considering the field star homogeneous abundances from Roederer et al. 2014,  we only take into account stars with abundances estimated
from 3 or more lines in order to select a high quality sample. We do not consider stars for which only upper limits were provided. The APOGEE sample is
composed of Gaia DR2 stars with parallax $>$ 0.3 mas, G $<$ 15~mag and Vtot$ >$ 180 km/s. 
Our final sample comprises 972 objects with APOGEE DR14 [Mg/Fe] abundances. 
In addition, the chemical abundances of Milky Way satellites have been analyzed using a compilation  with metallicities [Fe/H] $<$ -0.5~dex, obtained from the SAGA database \citep[][]{Saga}. We gather stars with Y, Eu and Mg abundance determinations, excluding those with only upper
limits, carbon-enriched stars (defined as [C/Fe] $<$ 0.9~dex if [Fe/H] $<$ -1.0~dex) and objects reported as binaries.

To better understand the [Y/Eu] behaviour, a separate study of [Eu/Fe] and [Y/Fe] abundance trends with [Mg/Fe],
for Milky Way satellites of different luminosities can be performed. Figure~A.1 shows the Mg abundance with respect to iron as a function of [Fe/H] for stars belonging to low mass satellites (Ursa Minor, Draco and Carina; square symbols) and to higher mass satellites (Fornax, Sculptor and Leo I; circles). 
A colour code on the [Eu/Fe] and [Y/Fe]  abundances is used for panels {$\it a$} and {$\it b$}, respectively. Stars showing high [Mg/Fe]
values present lower [Eu/Fe] abundances than those of similar metallicity with lower [Mg/Fe] values. As a consequence, stars with
[Eu/Fe] abundances lower than about 0.5~dex display low-[Mg/Fe] abundances only for metallicities higher than around -1.75~dex, suggesting a faster chemical evolution
of their parent systems. Conversely, at a given metallicity, higher [Mg/Fe] stars tend to have slightly higher [Y/Fe] values than lower [Mg/Fe] stars. 
This suggests that lower mass systems tend to present higher [Eu/Fe] enrichments and lightly lower  [Y/Fe] abundances than more massive ones,
conducting to higher [Y/Eu] ratios as shown in Figure~1.

\begin{figure*}[ht]
\centering
\begin{tabular}{c c}
\includegraphics[width=8cm,height=5.5cm]{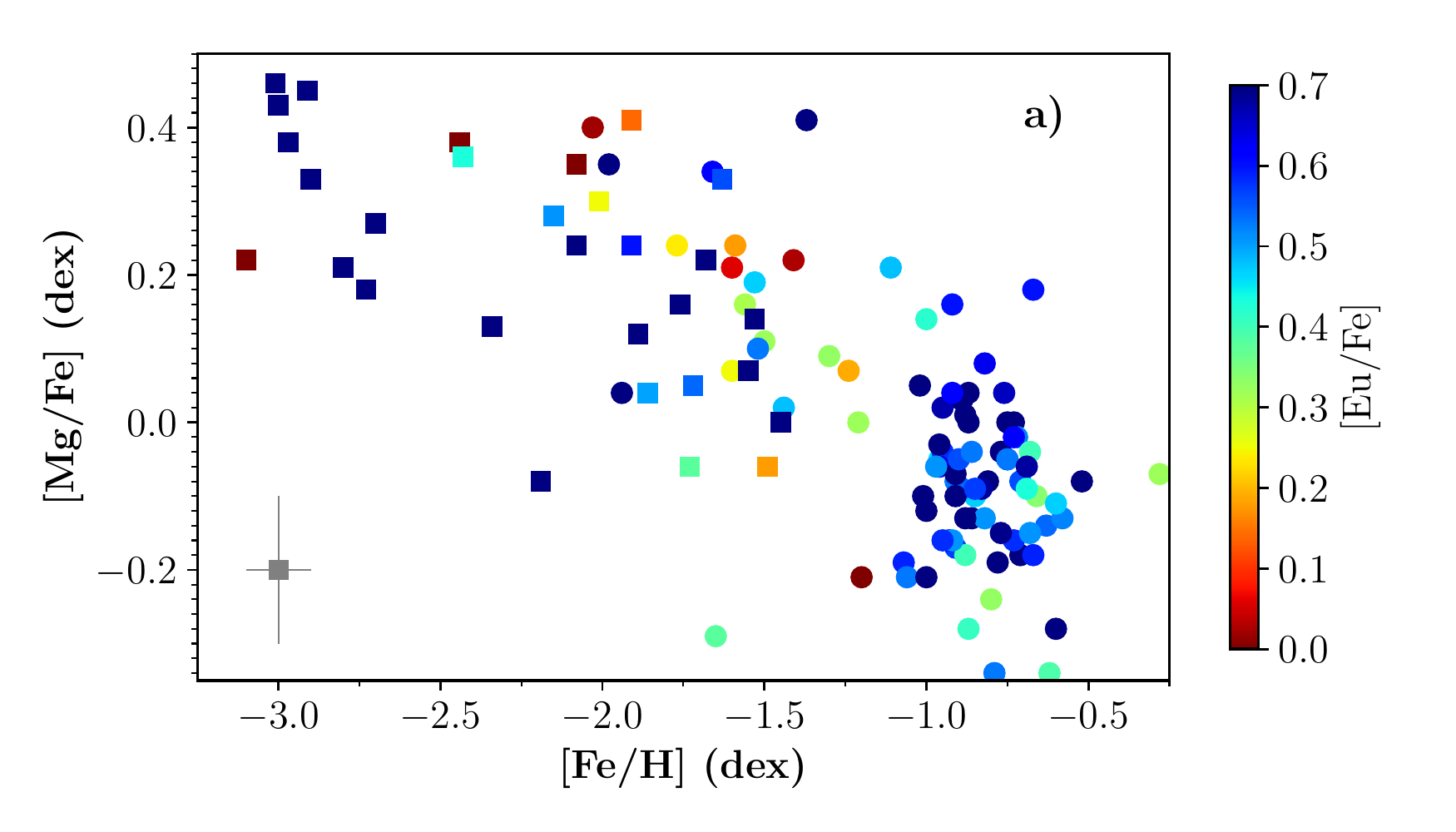} & \includegraphics[width=8cm,height=5.5cm]{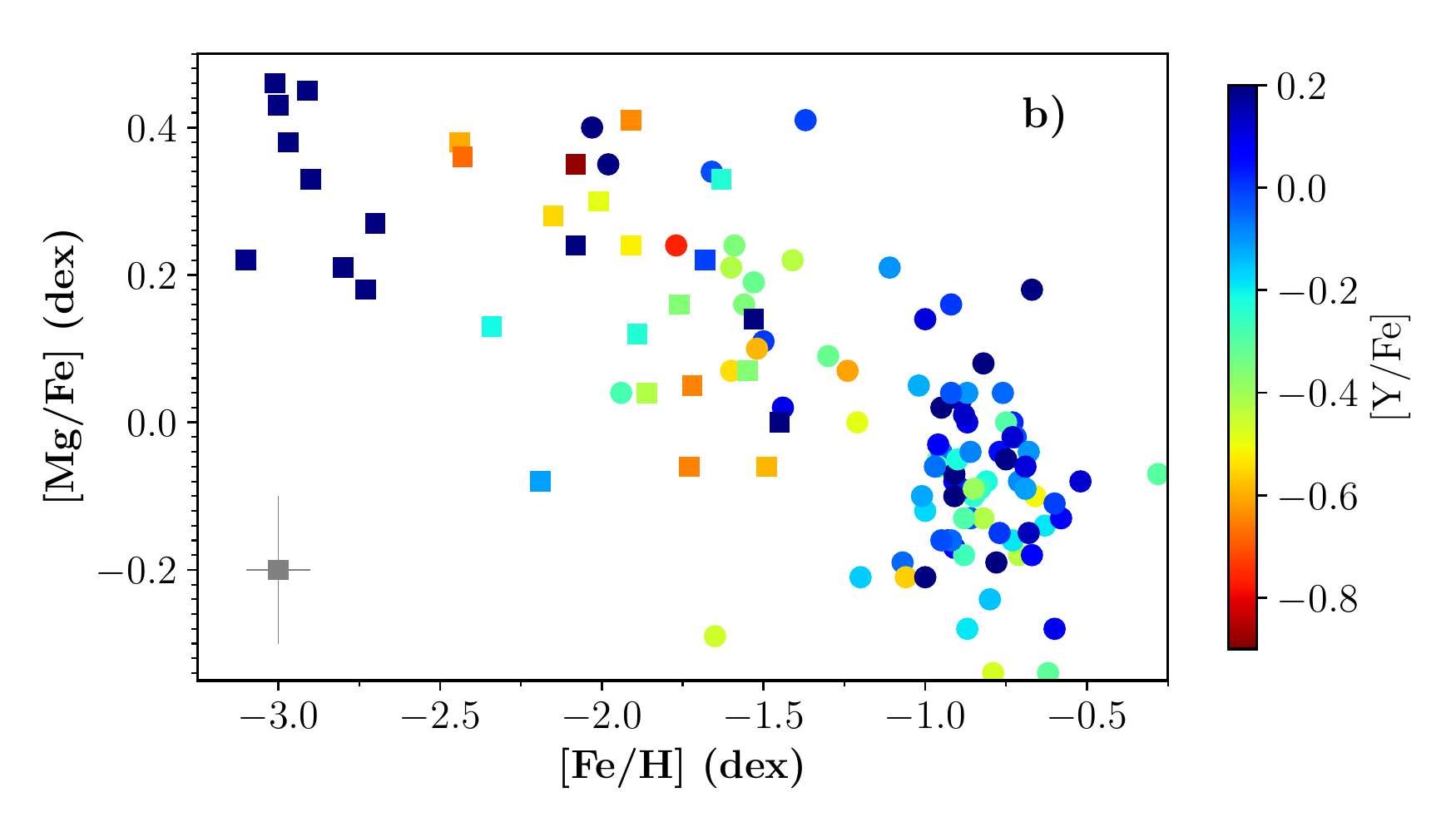} \\
\end{tabular}
\caption{Mg abundance with respect to iron as a function of [Fe/H] for stars belonging to
low mass satellites (Ursa Minor, Draco and Carina; square symbols) and to higher mass satellites (Fornax, Sculptor and Leo I; circles).
Points are colour coded by the stars [Eu/Fe] content (panel a) and by their [Y/Fe] abundance (panel b). 
}
\label{Inclin}
\end{figure*}

\begin{table*}[ht]
\caption{Adopted references for yttrium, europium and magnesium abundances}
\centering
\begin{tabular}{|l | c | c |}
\hline
\hline
 & &\\
\bf Population & \bf [Y/Fe] \&  [Eu/Fe] references & \bf [Mg/Fe] references\\
 & &\\
\hline
\hline
 & &\\
& \cite{Johnson17} \& references in their table 5, \cite{McWilliam2298},  &  \\ 
                Milky Way clusters                &   \cite{Shetrone03}, \cite{Munoz3201}, \cite{Roederer11} ,          & \cite{DualGalaxy18}\\
                               & \cite{Massari6362},    \cite{James6752} &\\
 & &\\                               
\hline
 & &\\
Milky Way field stars &  \cite{Fishlock17}, \cite{Roederer14} &  \cite{Nissen2010} \\
 &  & \cite{ApogeeDR14}\\
  & &\\
\hline
 & &\\
Satellites field stars &   \cite{Saga} &  \cite{Saga}\\
 & &\\
\hline
\hline
\end{tabular}
\end{table*}

\end{appendix}

\end{document}